\documentclass{article}
\usepackage{graphicx}
\usepackage{amsmath}
\usepackage{harvard}

\input{tcilatex}

\begin{document}

\title{Mermin's pentagram and Bell's theorem}
\author{P.K.Aravind \\
%EndAName
Physics Department, Worcester Polytechnic Institute, Worcester, MA 01609}
\date{\today}
\maketitle

\begin{abstract}
It is shown how a proof of the Bell-Kochen-Specker (BKS) theorem given by
Kernaghan and Peres can be experimentally realized using a scheme of
measurements derived from a related proof of the same theorem by Mermin. It
is also pointed out that if this BKS experiment is carried out independently
by two distant observers who repeatedly make measurements on a specially
correlated state of six qubits, it provides an inequality-free demonstration
of Bell's theorem as well.
\end{abstract}

Beginning with the groundbreaking work of Greenberger, Horne and Zeilinger
(GHZ)[1], the last decade has seen several new proofs of the Bell[2] and
Bell-Kochen-Specker(BKS)[3] theorems as well as a better understanding of
the relationship between these two fundamental theorems[4]. One interesting
line of work has focused on obtaining joint proofs of the BKS and Bell
theorems, the idea\textit{\ }being to first prove the BKS theorem and then
use a suitable strategem to convert this proof into a proof of Bell's
theorem.[5]. The three-particle GHZ state serves as the springboard for at
least three such joint proofs of the BKS and Bell theorems, as detailed in
the three scenarios below:

\textbf{Scenario 1}. Three qubits are given to three widely separated
observers, each of whom can measure two observables on his/her qubit. By
considering a set of ten observables pertaining to these qubits, Mermin[4]
gave a state-independent proof of the BKS theorem and then showed how to
convert it into a proof of Bell's theorem by assuming that the three qubits
were in a GHZ state.

\textbf{Scenario 2}. The starting point for this variation was provided by
Kernaghan and Peres[6], who extracted a set of 40 states from Mermin's ten
observables in Scenario 1 and used them to give a ''non-coloring'' proof of
the BKS theorem (so called because the proof works by showing that it is
impossible to assign the color red or green to each of the states in
accordance with a simple set of rules). We[5] reinterpreted the 40 states of
Kernaghan and Peres as those of a spin-7/2 particle (which also has an
eight-dimensional state space), and their proof as a proof of the BKS
theorem for such a particle; then, by considering a pair of spin-7/2
particles in a singlet state, we showed how this BKS proof could be
converted into a proof of Bell's theorem.

\textbf{Scenario 3}. Six qubits are shared, three to each, by two distant
observers. Elements of the Mermin and Kernaghan-Peres BKS proofs in
Scenarios 2 and 3 are combined to obtain a state-independent proof of the
BKS theorem, which is then promoted into a proof of Bell's theorem by
assuming that the six qubits are in a special entangled state.

The purpose of this paper is to give a detailed expos\'{e} of Scenario 3,
which has not been discussed before. This scenario is of interest for
several reasons. In the first place, it demonstrates that the BKS proof of
Kernaghan and Peres[6] is a state-independent proof and also shows how it
can be implemented in the laboratory (neither of these points was obvious
earlier). And, secondly, it shows how this BKS experiment can double as a
Bell experiment if it is performed independently by two observers using a
suitably correlated set of six qubits. The interest of this proof is further
enhanced by the fact that it is a member of a heirarchy of joint BKS-Bell
proofs extending upwards to larger numbers of qubits. The first member of
this heirarchy, which involves four qubits shared among two observers, was
presented by Cabello[7] and, in a somewhat different form, by the author[8].
This heirarchy of proofs casts a new light on some well known entangled
states and also has possible applications to quantum cryptography.

We now proceed to our main task, which is to present a joint BKS-Bell proof
based on Scenario 3. Figure 1 shows ten observables pertaining to a system
of three qubits arranged at the vertices of a pentagram. The pentagrammic
framework is due to Mermin[5], but the variables we have placed at its
vertices are those of Kernaghan and Peres[6]. The important facts about this
figure are the following: (1) each observable has eigenvalues $\pm 1$, (2)
the four observables lying along any edge of the pentagram constitute a
complete commuting set, and (3) the product of the observables (and hence
their eigenvalues) along any edge is $+1,$ with the exception of the
horizontal edge (labelled E5) for which the product is $-1$. In Mermin's BKS
proof, an observer (Alice) carries out a measurement of all the commuting
observables along any edge of the pentagram and finds a set of $+1$s and $-1$%
s (reflecting the eigenvalues of the measured observables) satisfying the
product constraint just mentioned. Her inability to attach a consistent set
of values to the ten observables in such a way that all the product
constraints are met furnishes a proof of the BKS theorem.

In the present scheme, we allow Alice to arrive at a different proof of the
BKS theorem (actually just the one found by Kernaghan and Peres) by
permitting her to carry out more general measurements than permitted in
Mermin's scheme. To be precise, we allow Alice to carry out ''hybrid''
measurements that jointly involve a pair of intersecting edges of the
pentagram. Let us denote the five edges of Mermin's pentagram by the symbols
E1 to E5, as indicated in Fig.1. Then each hybrid measurement can be
represented by a symbol such as Ex-Ey, indicating the two edges involved.
The hybrid measurement Ex-Ey is executed by first measuring the observable
at the intersection of edges Ex and Ey and following this with a measurement
of two more observables chosen from either Ex or Ey, depending upon the
outcome of the first measurement. The complete instructions for carrying out
a hybrid measurement can be summarized in a symbol of the form $(X|\left\{
Y_{1},Y_{2}\right\} \left\{ Z_{1},Z_{2}\right\} )$, which is more specific
than the vaguer (but still useful) symbol Ex-Ey. The precise instructions
for carrying out the measurement $(X|\left\{ Y_{1},Y_{2}\right\} \left\{
Z_{1},Z_{2}\right\} )$ are as follows: first measure the observable $X$ (at
the intersection of edges Ex and Ey) and follow this with a measurement of
the pair $Y_{1},Y_{2}$ if the eigenvalue of $X$ is $+1$ or the pair $%
Z_{1},Z_{2}$ if the eigenvalue of $X$ is $-1$. Note that the second pair of
observables measured always belongs to one of the edges involved, and that
the precise pair chosen is immaterial on account of the product constraint.

With the above notational conventions, we can present the 11 measurements
that Alice needs to be able to carry out on her qubits in order to validate
the BKS theorem. These measurements are listed, in both the alternative
notations introduced above, in the first column of Fig.2 (note that the last
measurement is actually a ''degenerate'' measurement involving only a single
edge of the pentagram). The eight mutually exclusive outcomes of each
measurement are listed immediately after it, with each outcome being
indicated in two ways: by the eigenvalues of the three observables measured
in producing it (this is done at the head of the column containing the
outcome) and by a number between 1 and 36. The reason the same number is
attached to several seemingly different outcomes (such as the fourth outcome
in the ninth row and the second outcome in the tenth row) is that these
outcomes are actually the same, as can be verified by supplying the
eigenvalue of the fourth, commuting observable to each of the outcomes and
comparing the complete lists of eigenvalues so obtained.

If one notes that each of the 36 outcomes occurs an even number of times
(either twice or four times) in Fig.2, the stage is set for a proof of the
BKS theorem. If Alice is a ''realist'' and believes that she only discovers
what already exists, she would be tempted to assign a definite value to each
of the outcomes (1 if it is preordained to occur or 0 if it is not) in such
a way that there is exactly one 1, and hence seven 0s, in each of the rows
of Fig.2. However this task is easily seen to be impossible as follows. For,
on the one hand, the total number of 1s in the table is required to be odd
(because there must be exactly one 1 in each row) while, on the other hand,
it is required to be even (because each outcome assigned the value 1 occurs
an even number of times). This contradiction discredits the assumption of
realism and proves the BKS theorem. However the damage to realism is not
fatal because of the assumption of noncontextuality made in assigning values
to the outcomes (by this it is meant that an outcome is assigned the same
value no matter which set of commuting observables is used to determine its
character). The assumption of noncontextuality has no empirical basis, and
so it, and the BKS theorem to which it leads, are both regarded as somewhat
questionable by most physicists.

We now show how to justify the above assumption of noncontextuality on the
basis of the principle of locality, and thus convert the above BKS proof
into a proof of Bell's theorem. We do this by enlisting a second
experimenter (Bob), giving him three qubits of his own, and allowing him to
do everything Alice can. The crucial trick needed to ensure that the
individual proofs of the BKS theorem arrived at by Alice and Bob can be
fused into their joint proof of Bell's theorem is that the six qubits
distributed to them are in the entangled state

\begin{equation}
\left| \Psi \right\rangle =\frac{1}{\sqrt{8}}\left[ 
\begin{array}{c}
\left| 000000\right\rangle +\left| 001001\right\rangle +\left|
010010\right\rangle +\left| 011011\right\rangle \\ 
+\left| 100100\right\rangle +\left| 101101\right\rangle +\left|
110110\right\rangle +\left| 111111\right\rangle
\end{array}
\right] ,
\end{equation}
where 0 and 1 refer to the basis states of a single qubit and it is
understood that the first three qubits are given to Alice and the last three
qubits to Bob. To forestall any confusion, we restate the complete
experiment to be carried out by Alice and Bob. The six-qubit state (1) is
generated (we will see below how) and the first three qubits are sent to
Alice and the last three to Bob. Alice and Bob each independently carry out
one of the 11 measurements in Fig.2 on their qubits, taking care to perform
their measurements outside each others light cones, and note the outcomes
they get. They then repeat this procedure as often as they like, using a
pristine entangled state (1) as the source of their qubits in each run.

The observations made individually by Alice and Bob confirm the BKS theorem,
for the reasons discussed earlier, while their combined observations suffice
to prove Bell's theorem if one notes the following remarkable correlations
between their observations in certain runs: in those runs in which the
hybrid measurements made by Alice and Bob have at least one edge of the
pentagram in common, the common outcomes associated with their measurements
either occur for both parties or are suppressed for both parties. For
example, if Alice and Bob both carry out the measurement E2-E5 and Alice
finds that outcome 24 occurs, Bob will find that this outcome occurs for him
too (note that the other seven common outcomes are suppressed for both
parties). As a second example, if Bob carries out the measurement E3-E4 and
Alice E4-E5 and Bob finds that neither outcome 9 nor 35 occurs, Alice will
find that neither of these outcomes materializes for her too. The
theoretical basis of these correlations is the following easily verified[9]
property of the state (1): if either person makes a measurement that
collapses his/her qubits into a state corresponding to one of the 40
outcomes of Fig.2, the other person's qubits collapse into this same state
too. One can use this correlation property to justify the assumption of
noncontextuality made by either observer in proving the BKS theorem and
thereby promote their individual BKS proofs into (their jointly achieved)
proofs of Bell's theorem. The argument that backs up this conclusion is
similar to that spelt out in our earlier proof[8] of Bell's theorem based on
four qubits. This completes the present joint proof of the BKS and Bell
theorems based on the six-qubit state (1).

The scheme of 11 measurements shown in Fig.2 and used in carrying out the
above proof is not unique. A careful examination of the properties of the 40
Kernaghan-Peres states (i.e. the simultaneous eigenstates of the five sets
of commuting observables in Fig.1) shows that there are exactly 320 distinct
schemes of such measurements[10].

The present BKS-Bell experiment is somewhat challenging to perform in the
laboratory because it requires each observer to carry out as many as three
measurements on a single qubit in a given run. Two strategies can be used to
carry out such measurements: one can use the technique of non-destructive
measurement which involves coupling the qubits to ancillas and making
measurements on the ancillas, or else one can generalize from two to three
measurements the technique discussed in ref [11].

The state (1) can be rewritten as

\begin{equation}
\left| \Psi \right\rangle =\frac{1}{\sqrt{2}}\left[ \left| 00\right\rangle
+\left| 11\right\rangle \right] _{14}\otimes \frac{1}{\sqrt{2}}\left[ \left|
00\right\rangle +\left| 11\right\rangle \right] _{25}\otimes \frac{1}{\sqrt{2%
}}\left[ \left| 00\right\rangle +\left| 11\right\rangle \right] _{36}
\end{equation}
from which it is obvious that it is a tensor product of three Bell states,
each involving one of Alice's qubits (subscripted 1,2 and 3) and one of
Bob's (subscripted 4,5 and 6) . Note that state (1) or (2) is quite
different from a six-particle GHZ state. It seems that GHZ-like states play
no role in the present proof, but that is not entirely true: if Alice or Bob
make certain measurements (e.g. E1-E5 or E5-E5) on their qubits, they might
end up with GHZ-like eigenstates associated with the edge E5.

It appears that the above proof can be generalized from 6 to $2n$ $(n>3)$
qubits using ideas put forward by Kernaghan and Peres[6]. However the
details of this generalization are not straightforward and cannot be
conveyed in a few sentences.

The present Bell experiment can be adapted to yield a system of quantum key
exchange, in a manner similar to that described earlier for our Bell proof
based on four qubits[8]. To do this, Alice and Bob carry out the Bell
experiment described above and then reveal publicly the measurements they
made in each of the runs. This allows them to establish a shared key whose
secrecy they can subsequently verify, or enhance, by sacrificing a portion
of the key. An interesting feature of this key is that it is an octal key,
based on an eight letter alphabet. Ternary[12] and quaternary[8] keys have
been discussed recently, suggesting some of the possibilities that lie
beyond binary (quantum) keys. It is not clear at present whether nonbinary
quantum keys are practically feasible and whether they offer any significant
advantages over binary keys. Nevertheless, it seems worth pointing out that
entangled states of qubits offer one route to the fabrication of such keys
and that they may be worth exploring further.

\bigskip

\ \ \ \ \ \ \ \ \ \ \ \ \ \ \ \ \ \ \ \ \ \ \ \ \ \ \ \ \ \ \ \ \ \ \ \ \ \
\ \ \ \ \ $\sigma _{x}^{3}$

\bigskip

\bigskip 

\ \ \ \ \ A \ \ \ \ \ \ \ \ \ \ \ \ \ \ \ \ \ \ \ \ \ \ \ B \ \ \ \ \ \ \ \
\ \ \ \ \ \ \ \ \ \ C \ \ \ \ \ \ \ \ \ \ \ \ \ \ \ \ \ \ \ \ \ \ \ \ \ D

\bigskip

\ \ \ \ \ \ \ \ \ \ \ \ \ \ \ \ \ \ \ \ \ \ \ \ \ \ \ \ \ \ \ \ \ \ \ \ \ \
\ \ \ \ \ \ \ \ \ \ \ \ \ 

\ \ \ \ \ \ \ \ \ \ \ \ \ \ \ \ \ \ \ \ \ \ \ \ $\sigma _{z}^{1}$ \ \ \ \ \
\ \ \ \ \ \ \ \ \ \ \ \ \ \ \ \ \ \ \ \ \ \ \ \ \ \ \ \ \ \ \ $\sigma
_{x}^{1}$

\ \ \ \ 

\ \ \ \ 

\ \ \ \ \ \ \ \ \ \ \ \ \ \ \ \ \ \ \ \ \ \ \ \ \ \ \ \ \ \ \ \ \ \ \ \ \ \
\ \ \ \ \ \ \ $\sigma _{z}^{3}$

\ \ \ \ \ \ \ 

\ \ \ \ \ \ \ \ \ \ \ \ \ \ \ \ \ \ \ \ \ \ \ \ \ \ \ \ \ \ \ \ 

\ \ \ \ \ \ \ \ \ \ \ \ \ \ $\sigma _{x}^{2}$ \ \ \ \ \ \ \ \ \ \ \ \ \ \ \
\ \ \ \ \ \ \ \ \ \ \ \ \ \ \ \ \ \ \ \ \ \ \ \ \ \ \ \ \ \ \ \ \ \ \ \ \ \
\ \ \ \ \ \ \ \ $\sigma _{z}^{2}$

\bigskip\ $\ \ \ \ \ \ \ \ \ \ \ \ \ \ \ \ \ \ \ \ \ \ \ \ \ \ \ \ \ \ \ \ \
\ \ \ \ \ \ \ \ \ \ \ \ \ \ \ \ \ \ \ \ \ \ \ \ \ \ \ \ \ \ \ \ \ \ \ \ \ \
\ \ \ \ \ \ \ \ \ \ \ \ \ \ \ \ \ \ \ \ \ \ \ \ \ \ \ \ \ \ \ \ \ \ \ $

\bigskip

\bigskip

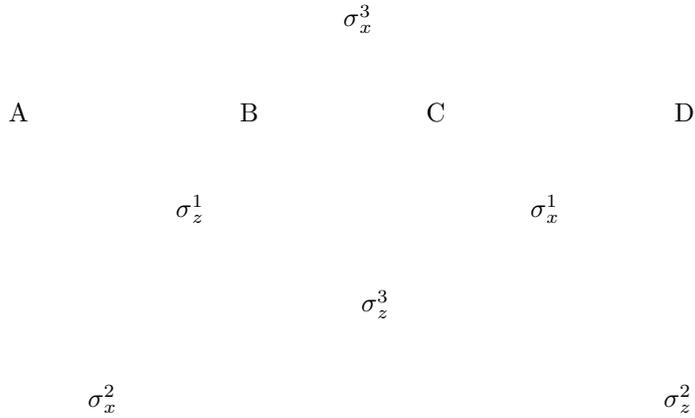
Fig.1. Mermin's pentagram, with ten observables pertaining to three qubits
placed at its vertices. The symbol $\sigma _{j}^{i}$ refers to the Pauli
operator $\sigma _{j}$ of particle $i$ and $A\equiv \sigma _{z}^{1}\sigma
_{z}^{2}\sigma _{z}^{3},B\equiv \sigma _{z}^{1}\sigma _{x}^{2}\sigma
_{x}^{3},C\equiv \sigma _{x}^{1}\sigma _{z}^{2}\sigma _{x}^{3}$ and $D\equiv
\sigma _{x}^{1}\sigma _{x}^{2}\sigma _{z}^{3}.$ The five edges of the
pentagram, labeled E1 to E5 (but not marked on the figure), each contain
four observables as follows: E1: $A,\sigma _{z}^{1},\sigma _{z}^{3},\sigma
_{z}^{2},$ E2: $\sigma _{x}^{3},B,$ $\sigma _{z}^{1},\sigma _{x}^{2},$ E3: $%
\sigma _{x}^{3},C,$ $\sigma _{x}^{1},\sigma _{z}^{2},$ E4: $D,\sigma
_{x}^{1},\sigma _{z}^{3},\sigma _{x}^{2},$ and E5: $A,B,C,D.$\newpage 

\ 
\begin{tabular}{|c|c|c|c|c|c|c|c|c|}
\hline
& $+++$ & $++-$ & $+-+$ & $+--$ & $-++$ & $-+-$ & $--+$ & $---$ \\ \hline
E1-E5 $(A|\{B,C\}\{\sigma _{z}^{1},\sigma _{z}^{2}\})$ & 1 & 2 & 3 & 4 & 5 & 
6 & 7 & 8 \\ \hline
E1-E4 $(\sigma _{z}^{3}|\{\sigma _{x}^{1},\sigma _{x}^{2}\}\{\sigma
_{z}^{1},\sigma _{z}^{2}\})$ & 9 & 10 & 11 & 12 & 5 & 13 & 14 & 8 \\ \hline
E1-E3 $(\sigma _{z}^{2}|\{\sigma _{x}^{1},\sigma _{x}^{3}\}\{\sigma
_{z}^{1},\sigma _{z}^{3}\})$ & 15 & 16 & 17 & 18 & 6 & 13 & 19 & 8 \\ \hline
E1-E2 $(\sigma _{z}^{1}|\{\sigma _{x}^{2},\sigma _{x}^{3}\}\{\sigma
_{z}^{2},\sigma _{z}^{3}\})$ & 20 & 21 & 22 & 23 & 7 & 14 & 19 & 8 \\ \hline
E2-E5 $(B|\{\sigma _{x}^{2},\sigma _{x}^{3}\}\{C,D\})$ & 20 & 24 & 25 & 23 & 
3 & 26 & 27 & 4 \\ \hline
E2-E4 $(\sigma _{x}^{2}|\{\sigma _{x}^{1},\sigma _{z}^{3}\}\{\sigma
_{z}^{1},\sigma _{x}^{3}\})$ & 9 & 28 & 11 & 29 & 22 & 23 & 25 & 30 \\ \hline
E2-E3 $(\sigma _{x}^{3}|\{\sigma _{x}^{1},\sigma _{z}^{2}\}\{\sigma
_{z}^{1},\sigma _{x}^{2}\})$ & 15 & 31 & 17 & 32 & 21 & 23 & 24 & 30 \\ 
\hline
E3-E5 $(C|\{A,B\}\{\sigma _{x}^{1},\sigma _{z}^{2}\})$ & 1 & 3 & 33 & 26 & 16
& 31 & 17 & 34 \\ \hline
E3-E4$(\sigma _{x}^{1}|\{\sigma _{x}^{2},\sigma _{z}^{3}\}\{\sigma
_{z}^{2},\sigma _{x}^{3}\})$ & 9 & 28 & 10 & 35 & 17 & 18 & 32 & 34 \\ \hline
E4-E5 $(D|\{\sigma _{x}^{1},\sigma _{x}^{2}\}\{A,B\})$ & 9 & 35 & 29 & 12 & 1
& 4 & 36 & 26 \\ \hline
E5-E5 $(A|\{B,C\}\{B,C\})$ & 1 & 2 & 3 & 4 & 33 & 36 & 26 & 27 \\ \hline
\end{tabular}

\bigskip

\bigskip

Fig.2. Each row of the above table shows a hybrid measurement based on
Mermin's pentagram, Fig.1, followed by its eight possible outcomes. Each
hybrid measurement is denoted by two alternative symbols: $Ex-Ey$, denoting
the pair of edges involved, and $(X|\{Y_{1},Y_{2}\}\{Z_{1},Z_{2}\}),$
containing more precise instructions on how to execute that measurement.The
measurement $(X|\{Y_{1},Y_{2}\}\{Z_{1},Z_{2}\})$ is carried out by first
measuring the observable $X$ at the intersection of edges $Ex$ and $Ey$ and
following this with a measurement of the pair $Y_{1},Y_{2}$ (if the
eigenvalue of $X$ is +1) or $Z_{1},Z_{2}$ (if the eigenvalue of $X$ is -1).
The effect of this measurement is to produce a collapse into an eigenstate
associated with edge $Ex$ or edge $Ey$. Each outcome (or eigenstate)
resulting from a hybrid measurement is indicated in two different ways: by
the eigenvalues of the observables measured in producing it (this is done at
the head of the column containing the outcome), and by a number between 1
and 36. Each outcome/eigenstate occurs either two or four times throughout
the entire table. This table gives a state-independent proof of the BKS
theorem for either observer as well as a state-dependent proof of Bell's
theorem achieved by both observers if they use the six-particle entangled
state (1). (Note: $A$ $\equiv $ $\sigma _{z}^{1}\sigma _{z}^{2}\sigma
_{z}^{3},B\equiv \sigma _{z}^{1}\sigma _{x}^{2}\sigma _{x}^{3},C\equiv
\sigma _{x}^{1}\sigma _{z}^{2}\sigma _{x}^{3}$ and $D\equiv \sigma
_{x}^{1}\sigma _{x}^{2}\sigma _{z}^{3}$ in the table above).

\bigskip


\begin{thebibliography}{99}
\bibitem{9}  This verification is easily carried out in two steps. First,
one notes that a measurement that collapses Alice's qubits into the state $%
a\left| 000\right\rangle +b\left| 001\right\rangle +c\left| 010\right\rangle
+d\left| 011\right\rangle +e\left| 100\right\rangle +f\left|
101\right\rangle +g\left| 110\right\rangle +h\left| 111\right\rangle $
collapses Bob's qubits into the same state but with the coefficients $%
a,b,c...$ replaced by their complex conjugates. Then one notes that each of
the forty Kernaghan-Peres states has real components in the standard Pauli
basis, implying that if Alice makes a measurement that causes her qubits to
collapse into one of these states, Bob's qubits collapse into the same state
too.
\bibitem{10}  The number of 320 can be understood as follows. The 40
Kernaghan-Peres states form 25 octads of eight mutually orthogonal states,
five of which might be termed ''pure octads'' and the remaining twenty as
''mixed octads''. Each pure octad consists of the eight simultaneous
eigenstates of one of the sets of commuting observables in Mermin's
pentagram. Each mixed octad consists of half the states of one pure octad
combined with half the states of another. There are ten ways in which the
pure octads can be paired up together, and each such pair gives rise to two
mixed octads that are the complements of each other (in the sense that they
have no members in common). If one deletes four mutually orthogonal states
from the 40 states, making sure that all the deleted states come from
distinct pure octads, one finds that there are exactly 11 octads not
involving any of the deleted states and having the properties required for
the BKS-Bell proof in Fig.2. A simple counting argument shows that the
number of ways in which four states can be deleted from the complete set of
forty in keeping with the required constraints is $5\cdot 8\cdot 4\cdot
2=320,$ which is a factor of 4 smaller than that reported in ref.6.
\end{thebibliography}
\end{document}